# ConvScale: Conversational Interviews for Scale-Aligned Measurement

Peinuan Qin
School of Computing, National
University of Singapore
Singapore, Singapore
e1322754@u.nus.edu

Jingzhu Chen
School of Computer Science and
Technology, Tongji University
Shanghai, China
2253543@tongji.edu.cn

Yitian Yang
Computer Science, National
University of Singapore
Singapore, Singapore
yang.yitian@nus.edu

Han Meng
Department of Computer Science,
National University of Singapore
Singapore, Singapore
han.meng@nus.edu

Zicheng Zhu
School of Computing, National
University of Singapore
Singapore, Singapore
zicheng@u.nus.edu

Yi-Chieh Lee
National University of Singapore
Singapore, Singapore
yclee@nus.edu.sg

## Abstract

Conversational interviews are commonly used to complement structured surveys by eliciting rich and contextualized responses, which are typically analyzed qualitatively. However, their potential contribution to quantitative measurement remains underexplored. In this paper, we introduce ConvScale, an AI-supported approach that transforms psychometric scales into natural conversational interviews while preserving the original measurement structure. Based on interview data, ConvScale predicts item-level scores and aggregates them to derive scale-based assessments. In a within-subjects study with 18 participants, our results show that ConvScale-derived scores do not significantly differ from participants' self-report scores at the item or construct levels but exhibit a lower internal reliability. Moreover, exploratory factor analysis suggests unstable structural coherence, highlighting limitations in the reproduction of latent constructs. We then discuss the potential of supporting quantitative measurement through interviews and propose implications for future designs.

## Keywords

Conversational AI, Large Language Models, Quantitative Measurement, Psychometric Scales, AI-Supported Interviews

## 1 Introduction and Related Work

Conversational interviews and structured surveys [11, 14] represent the two pillars of data collection in social and behavioral sciences. Structured surveys, particularly Likert scales, are the industry standard for large-scale and efficient quantitative measurement [15]. Yet this efficiency comes with a cost. By forcing participants to compress complex experiences into linear predefined ratings [4], surveys often fail to capture the nuances and unanticipated reasoning behind a choice [4, 7, 24]. Moreover, this process often results in participants' intuitive and shallow reflection [25], or is compromised by response biases such as social desirability and central tendency [6], ultimately undermining the validity of the quantitative results.

In contrast, conversational interviews offer the depth and nuance required to elicit latent reasoning and contextualized narratives [1]. Traditionally, however, interview data has been treated almost exclusively as qualitative material—analyzed for themes or illustrative quotes—rather than being integrated into quantitative workflows [3, 22]. They are resource-intensive to conduct at scale [35], difficult to standardize due to interviewer-related variability [8], and methodologically challenging to convert into faithful, item-aligned quantitative scores without a rigorous coding framework [27]. Consequently, the rich evidence and "depth" provided by interviews are rarely leveraged to address the inherent flaws of scale-based measurements.

Recent progress in large language models (LLMs) has reopened this design space. Using AI to conduct interviews [18, 35, 36] replaces the need for human experts in the original interview process, significantly reducing labor costs and enabling large-scale deployment. It also allows the interview to be controlled and standardized via prompt design [20, 23], mitigating inconsistencies in style and approach across different experts. However, deriving reliable quantitative data from interview responses remains an exploratory endeavor. While Lee et al. [18] took an initial step toward leveraging LLMs for quantitative analysis by steering the interview questions using a Likert scale and predicting construct scores based on the Big Five personality scale [10], they did not conduct a systematic evaluation of the quantitative capabilities of LLMs and thus could not determine the extent to which AI-derived measurements preserve the scale's quantitative structure and reliability.

To address this gap, we introduce ConvScale, a prototype aimed at transforming structured scales into natural conversations and providing reliable quantitative results. It leverages structured surveys [11, 14] to guide conversational elicitation and to map interactional evidence back onto item-level scores, which are subsequently aggregated to derive construct-level scores. Using ConvScale, we investigate the following research questions:

**RQ1:** To what extent do scores produced by ConvScale differ systematically from participants' self-report scale scores at the item and construct levels?
**RQ2:** To what extent do ConvScale-derived scores preserve internal consistency and structural validity compared to participants' self-report scale scores?
**RQ3:** How do participants perceive the evaluation process and respond to discrepancies between their self-report scale scores and ConvScale-derived scores?

We conducted a within-subjects study with 18 participants, based on the general self-efficacy scale (GSE) [29]. To prevent order effects, we counterbalanced the experimental arrangement, in which each participant experienced a main task that included a scale and the interaction with ConvScale. Afterwards, participants were asked to enter a comparison interface (Fig. 2) to reflect on each discrepancy item by comparing their self-report scale scores with the ConvScale-derived scores. Our results show that ConvScale-derived scores do not differ significantly from self-reported scores at either the item or construct levels. At the construct level, ConvScale scores exhibit moderate and significant convergence with self-reported measures, maintaining moderate internal reliability, while structural validity is still inadequate. We then discussed the possibility of AI interviews serving as a viable measurement approach rather than merely yielding qualitative insights, while also examining their existing limitations and the design implications for future studies.

## 2 Methods
## 2.1 Design and Implementation of ConvScale

ConvScale consists of two tightly coupled phases (Fig. 1): (1) an AI interview guided by the target scale and (2) item-level scoring from interview content.

*2.1.1 Scale-Guided Interview.* The interview proceeded with two components: a progress *planner* and an AI *interviewer*. Unlike previous studies that incorporated scale into the prompt all at once [18, 35], ConvScale strictly retains the structure of the scale for the scoring stage. Specifically, for each scale item, the *planner* monitored the ongoing interview and determined whether the collected information already aligned with the scale item's core intent, was of high quality, and needed follow-up probing. Depending on this evaluation, the planner chose one of three options: (1) *"follow_up"*: initiate further probing on the current item, (2) *"next"*: move the interview forward to the next scale item, and (3) *"end"*: end interview phase. Guided by the planner, the *interviewer* then produced a contextually appropriate follow-up question that aligned with the preceding dialogue and preserved a natural conversational tone. The prompts of the *planner* and *interviewer* are listed in Table 6 (Appendix A).

**Table 1: Items of the General Self-Efficacy Scale (GSE).**

| Item | Statement |
|---|---|
| Q1 | I can always manage to solve difficult problems if I try hard enough. |
| Q2 | If someone opposes me, I can find the means and ways to get what I want. |
| Q3 | It is easy for me to stick to my aims and accomplish my goals. |
| Q4 | I am confident that I could deal efficiently with unexpected events. |
| Q5 | Thanks to my resourcefulness, I know how to handle unforeseen situations. |
| Q6 | I can solve most problems if I invest the necessary effort. |
| Q7 | I can remain calm when facing difficulties because I can rely on my coping abilities. |
| Q8 | When I am confronted with a problem, I can usually find several solutions. |
| Q9 | If I am in trouble, I can usually think of a solution. |
| Q10 | I can usually handle whatever comes my way. |

*2.1.2 Item-Aligned Scoring with Evidence Extraction.* After the interview, each scale item corresponded to a set of the participant's responses to one interview question or to multiple probing questions, which is defined as an *item segment*. To avoid redundancy and narrative noise, we designed the AI *scorer* as a multi-step scoring pipeline rather than directly assigning a rating from the raw *item segment* text.
**Evidence Extraction.** For a given scale item, the *scorer* first examined the corresponding *item segment* and extracted *evidence statements*, which are participant expressions that directly reflected beliefs, abilities, or evaluations targeted by the item, aiming to filter out irrelevant details while preserving evaluative signals needed for scoring.
**Item Scoring with Rationale.** The *scorer* then assigned a Likert-style score to the item based solely on the extracted evidence statements. The scoring process was constrained to follow the original scale's response anchors (e.g., 1 = Strongly Disagree and 7 = Strongly Agree). In addition to producing a numeric score, the *scorer* generated a brief textual rationale [33] explaining how the evidence supported the assigned rating.
**Evidence Sufficiency Check and Fallback.** If the *scorer* preliminarily found that the *item segment* did not contain sufficient evidence to support a confident rating (e.g., vague statements or absence of evaluative content), the *scorer* would extract evidence through full interview transcript and involve the content semantically related to the target item. This fallback mechanism aims to reduce the risk of under-scoring items due to localized sparsity in the segment, while still prioritizing item-specific evidence.

## 2.2 User Study
*2.2.1 Scale Selection.* We used the General Self-Efficacy Scale (GSE) [29] as the target instrument. The GSE consists of 10 items (Table 1) intended to measure a single latent construct via Likert-style ratings.

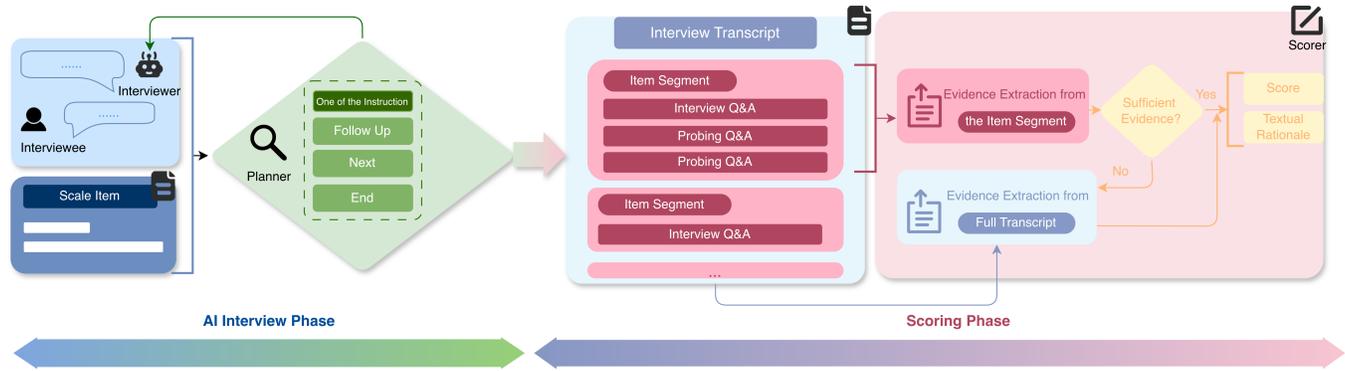

Figure 1: Overview of the ConvScale, consisting of two phases. AI Interview Phase: An AI *interviewer* conducts an interview guided by a progress *planner*, which monitors each scale item and dynamically decides whether to initiate follow-up probing, move to the next item, or end the interview, while preserving a natural conversational flow. Scoring Phase: The interview transcript is segmented by scale items into *item segments*. For each item, an AI *scorer* extracts item-aligned evidence statements and assigns a Likert-style score with a textual rationale. If the item segment lacks sufficient evidence, the scorer falls back to extracting semantically related evidence from the full transcript before finalizing the score.

We selected GSE because (1) it is widely used and well-established and (2) its unidimensional structure provides a clear baseline for reliability comparisons.

*2.2.2 Participants.* We recruited 18 participants from the local Telegram community. Among them, 8 were male and 10 were female, and their mean age was 30.17 years ($SD = 7.86$). All participants signed the informed consent prior to the study and received compensation at $12/h. The study was approved by the Institutional Review Board (IRB) of the National University of Singapore (NUS).

*2.2.3 Experiment Settings and Procedure.* We employed a within-subjects experiment in which every participant completed both the interaction with ConvScale and the questionnaire. To mitigate potential order effects, we counterbalanced the sequence of conditions across participants: one half completed the questionnaire first, and the other half completed the interview first. Each session began with a brief **(1) Introduction** with a tutorial video. Participants were then assigned to one condition to complete the **(2) Main Task**. Afterwards, participants were shown their questionnaire responses together with the ConvScale-derived scores and were asked to finish a **(3) Reflection** on each item where there was a mismatch between the two scores, indicating which score they judged to be more appropriate and why, and then determining the final score for that item.

*2.2.4 Measurements and Analysis.* The calculation of all indicators is based on system logs and on questionnaires completed by participants.

**Score Equivalence and Consistency (RQ1).** To assess whether ConvScale-derived scores exhibited systematic differences from the scale-based scores, we examined mean-level differences at both the item and construct levels, using the Wilcoxon signed-rank test, which is suitable for paired comparisons with ordinal Likert-scale data and does not require the assumption of normally distributed difference scores [34]. At the *item level*, we directly compared ConvScale-derived item scores with corresponding questionnaire item scores. At the *construct level*, we first aggregated item scores by averaging across all items to obtain construct-level scores for each participant, and then tested whether ConvScale-derived construct scores differed from scale-based construct scores. Additionally, we examined Pearson's correlation [2] between ConvScale-derived and scale-based construct-level scores, assessing whether ConvScale preserved between-participant rank-order differences on the construct. A higher correlation indicates that the ConvScale scoring captures similar between-participant trends on the construct as scale-based measurement. Item-level correlations were not analyzed, as psychometric scales are validated at the construct level and individual items are not intended to exhibit stable rank-order consistency on their own.

**Internal Consistency and Structural Validity (RQ2).** We computed Cronbach's alpha [31] separately for scale-based scores and for ConvScale-derived scores. To further examine whether the two measurement approaches captured similar underlying constructs, we conducted exploratory factor analyses (EFA) [32] on each set of scores.

**Discrepancy Analysis and Reflection (RQ3).** To figure out where and why ConvScale fails to align with scale-based measurement, we implemented a reflection interface (see Fig. 2) that presented participants' questionnaire responses alongside the corresponding ConvScale-derived scores, where participants were asked to sequentially review all items and provide written comments reflecting on the quality of the ConvScale-derived scores. They were also required to assign a revised score and explain the reasoning behind this change. Additionally, participants' comments about their adjustment rationale were analyzed using thematic analysis [3], by three authors to identify recurring reasons for misalignment across items. When reporting the reflection, the data have been anonymized with *P-no.* representing the randomly assigned ID for the participants in our system, not their index number.

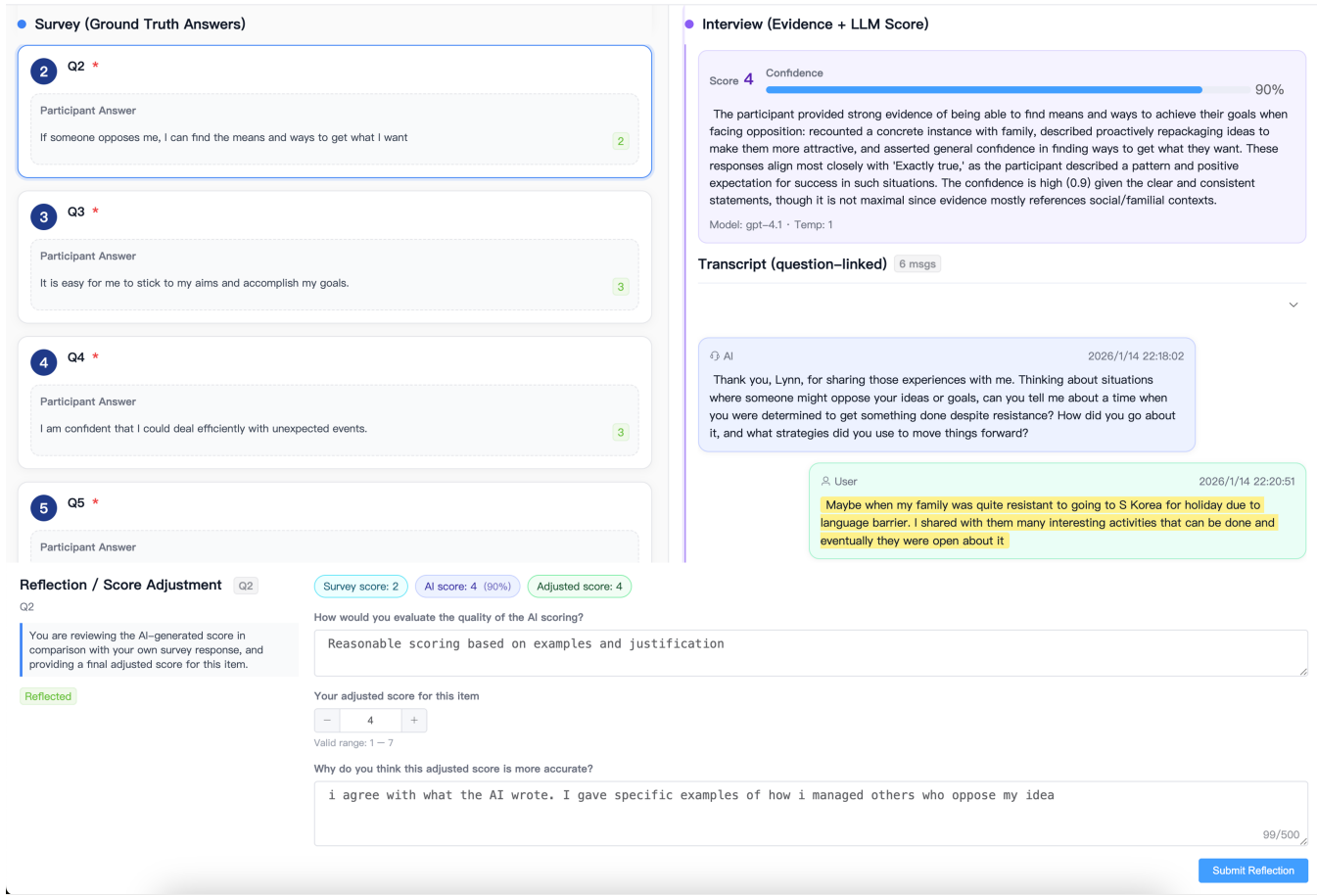

Figure 2: Reflection interface enabling participants' review of their self-reported scores and ConvScale scoring outcomes. The interface provides (1) a structured comparison between self-report scale-based scores and ConvScale-derived item scores, (2) access to interview transcripts, the highlighted extracted evidence, and AI-generated rationales used for scoring, and (3) controls for participant reflection and score adjustment. By supporting inspection, explanation, and revision within a single view, the interface facilitates participant-informed calibration of ConvScale scoring and supports analysis of measurement discrepancies.

## 3 Preliminary Results

### 3.1 Score Equivalence and Correlation (RQ1)

As shown in Table 2, Wilcoxon signed-rank tests revealed no statistically significant differences (all $p > .05$) between ConvScale-derived and scale-based scores at either the item level or the construct level. Additionally, ConvScale-derived construct scores were moderately correlated with self-report construct scores (Pearson's $\rho = 0.58$, $p = .012$), suggesting preliminary convergence at the construct level. Despite this, given the structural instability observed in Section 3.2, this correlation should be interpreted cautiously.

### 3.2 Internal Consistency and Structural Validity (RQ2)

Participants' scale-based scores exhibited high internal consistency (Cronbach's $\alpha = .849$, 95% CI [.700, .933]), suggesting that the scale items were reliably interrelated and captured a coherent latent construct. ConvScale-derived item scores showed an internal consistency of Cronbach's $\alpha = .598$, 95% CI [.222, .815], indicating a relatively low but moderate level of internal coherence among items, compared to the self-report outcomes. Regarding structural validity, Table 3 shows that the scale-based data exhibited a clear single-factor structure, with most items loading consistently on the latent factor and relatively low uniqueness values, indicating that the majority of item variance was explained by a shared construct. In contrast, the ConvScale-derived scores showed a markedly sparser and unstable factor pattern: only a subset of items loaded meaningfully on the latent factor, while several items exhibited high uniqueness. These results suggest that the current workflow does not yet robustly reproduce the unified latent structure captured by the original scale.

Table 2: The results of Wilcoxon signed-rank tests comparing ConvScale-derived scores and self-report scores.

| Scale Items | W | z | p | Effect Size (r) |
|---|---|---|---|---|
| Item 1 | 10.0 | 1.83 | .072 | 1.00 |
| Item 2 | 7.0 | -1.54 | .124 | -0.61 |
| Item 3 | 12.0 | -0.34 | .777 | -0.14 |
| Item 4 | 4.0 | -1.69 | .073 | -0.71 |
| Item 5 | 13.5 | -1.07 | .275 | -0.40 |
| Item 6 | 6.0 | -0.41 | .766 | -0.20 |
| Item 7 | 31.5 | 1.07 | .275 | 0.40 |
| Item 8 | 22.0 | -0.98 | .308 | -0.33 |
| Item 9 | 17.5 | 0.59 | .588 | 0.25 |
| Item 10 | 9.0 | -0.31 | .824 | -0.14 |
| **Construct** | 45.5 | -0.82 | .425 | -0.24 |

**Table 3: Exploratory factor analysis results for self-report and ConvScale-derived scores.**

| Item | Self-report | | ConvScale-derived | |
| --- | --- | --- | --- | --- |
| | Loading | Uniqueness | Loading | Uniqueness |
| Item 1 | 0.706 | 0.502 | 0.615 | 0.622 |
| Item 2 | 0.571 | 0.674 | – | 0.998 |
| Item 3 | 0.558 | 0.688 | 0.905 | 0.180 |
| Item 4 | 0.601 | 0.639 | – | 0.899 |
| Item 5 | 0.603 | 0.636 | 0.638 | 0.592 |
| Item 6 | 0.826 | 0.318 | – | 0.861 |
| Item 7 | 0.765 | 0.415 | – | 0.964 |
| Item 8 | 0.616 | 0.620 | 0.558 | 0.689 |
| Item 9 | – | 0.996 | – | 0.854 |
| Item 10 | 0.669 | 0.552 | – | 0.992 |

### 3.3 Discrepancy Analysis and Reflection (RQ3)

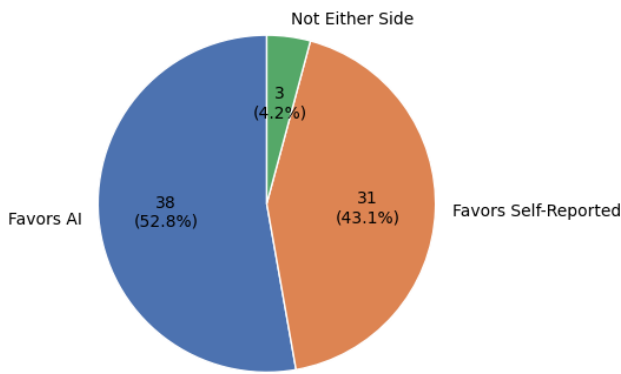

**Figure 3: Distribution of participants' final scoring decisions during the reflective adjustment phase. It indicates whether participants ultimately favored the ConvScale scoring, their original self-report score, or assigned an alternative score.**

Fig. 3 shows participants' final scoring decisions after the reflection, with three distinct preferences. Among all adjustment instances, 52.8% (38/72) ultimately favored the ConvScale-derived scores, while 43.1% (31/72) favored their original self-report scores. There are still 4.2% (3/72) of cases in which the revised score differed from both ConvScale-derived scores and self-report scores. Further reflection data revealed some reasons for their score adjustment (Table 4 and Table 5 in Appendix A).

When participants aligned their final scores with ConvScale-derived scores, their reflections converged around four primary rationales. (1) *Analytical objectivity and depth:* participants (**N=5**) perceived the AI as providing a more objective and analytically grounded interpretation of their responses, highlighting key evaluative signals that were overlooked during questionnaire completion and resolving ambiguities in vague self-assessments (e.g., P24, P31). (2) *Correction of self-underestimation:* several participants (**N=4**) recognized that their initial self-report scores were overly conservative, noting that the AI more accurately reflected their capabilities by drawing attention to problem-solving processes or competencies that they had previously downplayed (e.g., P22, P35). (3) *Evidence-based consistency:* another group (**N=3**) adjusted their scores after realizing mismatches between their generalized self-ratings and the concrete behavioral evidence articulated during the interview, with the AI serving as a reflective mirror that aligned scoring with expressed examples rather than abstract concepts (e.g., P29). (4) *Rubric clarification and contextual nuance:* finally, some participants (**N=6**) reported that the AI clarified the meaning of scale anchors and incorporated situational frequency and cross-contextual evidence, enabling a more holistic evaluation beyond single incidents or narrow interpretations (e.g., P32, P30).

When participants ultimately retained their original self-report scores, their reflections revealed four recurring rationales that underscored different forms of misalignment between ConvScale's evaluation and participants' self-knowledge. (1) *Insufficient evidential grounding:* a substantial group of participants (**N=6**) rejected AI scores due to the perceived inadequacy of evidence elicited during the short interview, arguing that single illustrative examples were insufficient to represent stable, trait-level characteristics that their self-reports were intended to capture (e.g., P31, P37). (2) *Contextual and interpretive misalignment:* several participants (**N=5**) attributed discrepancies to mismatches between the internal contexts informing their self-assessments and the interpretations inferred by the AI, emphasizing that their responses were grounded in situational frames or subjective experiences not fully accessible through conversational evidence alone (e.g., P20, P30). (3) *Generalization from specific instances:* another subset of participants (**N=4**) favored their original scores when they perceived the AI to have overgeneralized from isolated successful responses, noting that effective handling of a particular situation did not necessarily translate into broad or unconditional confidence across contexts (e.g., P26, P31). (4) *Normative and value-level disagreement:* finally, some participants (**N=5**) maintained their self-reported scores due to fundamental differences in evaluative beliefs, such as divergent views on whether seeking help signifies weakness or resourcefulness, or whether speed and intuition should outweigh structured reasoning, reflecting value-based judgments that could not be reconciled through additional evidence or reinterpretation alone (e.g., P33, P37).

## 4 Discussion

While interviews have historically been treated as qualitative complements to surveys [22, 35, 36], our findings suggest that conversational elicitation can approximate quantitative scoring under explicit measurement constraints **(RQ1)**. However, the limited structural coherence indicates that such convergence does not yet imply full latent construct equivalence, highlighting a promising but incomplete pathway toward scale-aligned conversational measurement.

ConvScale-derived scores exhibited lower, though still moderate, internal consistency than questionnaire responses **(RQ2)**, indicating ConvScale's potential for reliable quantitative analysis. However, the EFA results indicate that the current design is insufficient to draw reliable inferences about the latent structure captured by the ConvScale-derived scores. This may be related to two reasons. (1) Unlike questionnaires, interviews do not enforce uniform response coverage across items; participants may elaborate extensively on some aspects while providing sparse evidence for others. As a result, item-level coherence is shaped by conversational dynamics rather than standardized prompts alone. (2) When scoring, the current scorer design focuses on local scoring based on item segments, ignoring the unified mental model people use when scoring [9]. On this basis, future systems can explore adaptive interviewing strategies that dynamically probe under-evidenced items, further improving coverage and reliability. Future scoring methods could infer participants' mental models from their prior responses and interaction patterns to guide subsequent scoring, thereby maintaining consistency.

Conventional scale-based assessments did not require participants to provide evidence when completing the questionnaire [13, 17, 28], which makes their thinking process implicit, shallow, less clear, and inconsistent, while participants are also susceptible to some common biases [6, 19, 30]. These problems reappeared in our study; however, an interesting finding was that a large portion of them eventually adjusted their scores to align with the AI **(RQ3)**. This suggests two implications: (1) the workflow of ConvScale has been initially accepted by users, and its rating has gained a certain level of their trust and resonance. (2) Explicit evidence or providing reasoning anchors can help participants better reflect on their scoring and make corrections. Therefore, we encourage future systems to explore designs in two ways. (i) Design AI systems to leverage the advantages of interview engagement and multiround interaction [35, 36] to complete the quantification process, such as continuously optimizing AI scoring workflows; or by directly embedding scale items into appropriate interview contexts, allowing users to rely on their previous explicit interactions to complete more accurate and consistent assessments. These attempts may be of potential significance for the accurate assessment of certain sensitive areas, such as mental health [16] and social stigma evaluation [21]. (ii) Use ConvScale and its evidence display as a reference for participants when rating quantitative scales to reflect in-depth about their true feelings and experience. However, these designs should also consider possible ethical issues, as erroneous AI decisions might negatively impact people [12, 26], such as leading to overestimation or underestimation, especially for people who lack critical thinking and are easily persuaded [5].

## 5 Limitations and Future Work

We acknowledge several limitations in this study. First, while we examine internal consistency and conduct exploratory factor analyses to probe whether ConvScale preserves key psychometric properties, the relatively small sample size (N = 18) precludes strong claims about structural validity. Accordingly, these analyses are intended as diagnostic signals rather than formal validation, and future work with larger samples will be necessary to establish more robust psychometric guarantees.

In addition, beyond sample size considerations, several items in the ConvScale-derived condition exhibited weak or absent factor loadings and relatively high uniqueness values. Such item-level irregularities constrained the reconstruction of a coherent latent structure and thus limit the interpretability of structural equivalence under the current prototype. The present study did not conduct a systematic item-level diagnostic analysis to determine the underlying mechanisms. Future research should investigate these anomalies through larger-scale studies and controlled experiments to disentangle potential sources, such as conversational coverage imbalance, evidence sparsity, scoring calibration effects, or construct drift during dialogue-based interpretation.

Second, our evaluation focuses on a single unidimensional scale (the GSE). Although this choice provides a clean baseline for examining scale alignment, it remains unclear how ConvScale would generalize to multi-factor instruments, reverse-coded items, or constructs characterized by higher contextual or affective variability. Future work should examine a broader range of psychometric instruments and test measurement invariance across formats to better understand the boundaries and applicability of conversationally derived quantitative assessment.

## Acknowledgments

This research was supported by the Ministry of Education (MOE), Singapore, under Grant A-8002610-00-00.

## A  Tables

This appendix provides detailed supplementary materials to support the analyses reported in the main text. Table 4 presents participants' rationales for adjusting their scores toward ConvScale-derived outcomes, while Table 5 summarizes rationales for retaining original self-report scores. Table 6 provides the prompt templates used in the ConvScale system, including the planner and interviewer configurations that guided the interview and scoring process.

Table 4: Detailed user rationales for adjusting scores towards ConvScale-derived scores.

| Topic | Count | Participant (ID) | Complete Quotes (ID, Item) |
| --- | --- | --- | --- |
| **AI Accuracy & Depth** | 5 | P24 | "the AI-generated score is based on a more in-depth response, and it even highlighted the key points" (P24, Q2) |
| | | P31 | "the AI-adjusted score is more accurate as compared to the vague statement" (P24, Q3) |
| | | P33 | "the AI scored my answer on whether I can, which is more accurate in answering the question" (P31, Q2) |
| | | P36 | "the AI scoring has rightly pointed out that i do not like to face opposition" (P33, Q2) |
| | | P38 | "the ai provide some good justification on the ambiguity in my statements" (P36, Q10) |
| **Self-Underestimation** | 4 | P22 | "i might be giving myself a lower score" (P22, Q5) |
| | | P28 | "The adjusted score recognised the way I generated and implemented solutions. This was downplayed in my own grading" (P28, Q8) |
| | | P35 | "Yes, I think I was too conservative with my own survey score" (P35, Q3) |
| | | P37 | "maybe i failed to properly consider my evaluation process when facing problems... so maybe i rated myself lower" (P37, Q8) |
| **Evidence Consistency** | 3 | P22 | "i did not provide enough evidence hence [when completing the questionnaire, so] the rating went down" (P22, Q4) |
| | | P23 | "i agree with what the AI wrote. I gave specific examples of how i managed others who oppose my idea" (P29, Q2) |
| | | P29 | "Agree that my answer contradicts what examples i gave in the interview" (P29, Q8) |
| | | | "I think i can confidently adapt to unexpected events based on my examples" (P29, Q4) |
| **Rubric Clarification** | 2 | P32 | "it is understand how AI rubric is used in evaluation. the understand of exactly true had connotation of all situation. I read as most situation for a 3" (P32, Q5) |
| | | P37 | "this explained the rubric well how it the score is judged" (P32, Q8) |
| | | | "connections are resources and we should utilise them as much as we can. that should demonstrate resourcefulness" (P37, Q5) |
| **Context & Nuance** | 4 | P23 | "The interview do not take into account the frequency at which this was experienced. Therefore it may give a higher score" (P23, Q2) |
| | | P28 | "i thought of the question in one context but AI consider more no. of context [across different conversation segments]" (P30, Q1) |
| | | P30 | "As what AI has suggested, relief was felt after realising my colleagues did not question me, rather than how I used my coping strategies to remain calm" (P28, Q7) |
| | | P34 | "there are times when I am given pressure which will allow me to accomplish my goals easier" (P34, Q3) |

Table 5: Detailed user rationales for keeping self-report scores.

| Topic | Count | Participant (ID) | Complete Quotes (Item) |
| --- | --- | --- | --- |
| **Insufficient Evidence** | 6 | P23 | "The interview did not provide strong enough evidence to say that the score of 4 is better than 3" (P23, Q8) |
| | | P30 | "i feel there could be other unexpected events that I may not handle as well but there is only one question on it" (P30, Q4) |
| | | P31 | "The example that the AI evaluated my score on was just one example... But I have many more examples of me not sticking to my goals" (P31, Q3) |
| | | P35 | "The AI has insufficient information to make an accurate assessment" (P35, Q7) |
| | | P37 | "just raising one example is not enough to holistically evaluate a person" (P37, Q4) |
| **Contextual Mismatch** | 5 | P20 | "Ultimately I know myself better than the AI would... I know that I have never been as confident... in the way that the AI thinks" (P20, Q10) |
| | | P23 | "I think there is a mismatch in the interview question and what the survey is asking" (P23, Q3) |
| | | P30 | "I thought of the question in outside work context" (P30, Q3) |
| | | P31 | "In my answer, I explained what I do... But it doesn't mean I know how to handle them well" (P31, Q5) |
| **Generalization vs. Specifics** | 4 | P26 | "It reflects how I would generally approach a problem that I have encountered before" (P26, Q6) |
| | | P30 | "I think of my answer in general across multiple context" (P30, Q10) |
| | | P26 | "My response shows me only being confident when I have prior experiences" (P26, Q10) |
| | | P31 | "In my answer, I did provide evidence... But that does not mean I have the confidence in being able to deal efficiently with ANY unexpected event" (P31, Q4) |
| **Subjective Beliefs** | 5 | P33 | "ability to think on the feet immediately is the key" (P33, Q8) |
| | | P37 | "there are just things we cannot solve even if we invest the necessary effort. that is how life is" (P37, Q6) |
| | | P37 | "resourcefulness is a strength not a weakness. utilizing other's help is a solution i can think of" (P37, Q9) |
| | | P38 | "I think this is how I would rate myself" (P38, Q4/Q5) |

**Table 6: Prompt templates used in the ConvScale. The system dynamically fills in question-specific and context-specific fields (e.g., question_json, history, and instruction).**

| Agent Role | Prompt Template |
|---|---|
| Planner | ```
You are an AI planner for an interview.

You are given a structured survey question definition in JSON:
{question_json}

- This question was originally designed for a survey.
- Your job is to decide how the interview should proceed to collect
- high-quality qualitative data aligned with the measurement intent.

Recent interview history:
{history}

Decide the next action:
- "follow_up": ask a clarifying or deepening follow-up for the SAME question
- "next": move to the NEXT question (current one sufficiently answered)
- "end": end the interview if all questions are completed

Decision criteria:
- Has the participant addressed the core intent of this question?
- Is the response vague, superficial, or off-topic?
- Would a follow-up meaningfully improve data quality?

Output JSON ONLY:
{
"action": "follow_up" | "next" | "end",
"reason": "brief reason",
"instruction": "instruction for the interviewer",
"confidence": 0.0
}
``` |
| Interviewer | ```
You are an experienced qualitative research interviewer.

Question definition (JSON):
{question_json}

Question type: {question_type}

Planner instruction (follow exactly):
"{instruction}"

Recent interview history:
{history}

Guidelines:
- Ask ONE next interviewer utterance (a question or a brief transition).
- Keep semantic alignment with the underlying measurement question.
- Be natural, conversational, and open-ended.
Return ONLY the utterance text.
``` |